\documentclass[aps,prl,longbibliography,amsmath,amssymb,groupedaddress,superscriptaddress,floatfix,twocolumn,showkeys]{revtex4-1}
    \usepackage{graphicx}
    \usepackage{dcolumn}
    \usepackage{bm}
    \usepackage{color}
    \usepackage{xcolor}
    \usepackage{multirow}
    \usepackage{siunitx}
    \usepackage{ulem}
    \DeclareSIUnit{\sqrthz}{\ensuremath{\sqrt{\text{\hertz}}}}
    \usepackage{soul}
    \usepackage{makecell}
   
    \newcommand{\xzp}{x_{\text{zp}}} 
    
    \newcommand{\red}[1]{{\color{red}#1}}

    \newcommand{\minus}{\scalebox{0.65}[1.0]{$-$}}

\begin{document}
    
    \title{Analysis of membrane phononic crystals with wide bandgaps and low-mass defects}
    
    \author{C. Reetz}
        \address{JILA, National Institute of Standards and Technology and University of Colorado, and
    Department of Physics, University of Colorado, Boulder, Colorado 80309, USA}
    \author{R. Fischer}
    \address{JILA, National Institute of Standards and Technology and University of Colorado, and
    Department of Physics, University of Colorado, Boulder, Colorado 80309, USA}
    \address{Rafael Ltd, Haifa 31021, Israel}
    \author{G. G. T. Assump\c{c}\~{a}o}
        \address{JILA, National Institute of Standards and Technology and University of Colorado, and
    Department of Physics, University of Colorado, Boulder, Colorado 80309, USA}
    \author{D. P. McNally}
        \address{JILA, National Institute of Standards and Technology and University of Colorado, and
    Department of Physics, University of Colorado, Boulder, Colorado 80309, USA}
    \author{P. S. Burns}
        \address{JILA, National Institute of Standards and Technology and University of Colorado, and
    Department of Physics, University of Colorado, Boulder, Colorado 80309, USA}
    \author{J. C. Sankey}
    \address{Department of Physics, McGill University, Montreal, Canada}
    \author{C. A. Regal}
    \address{JILA, National Institute of Standards and Technology and University of Colorado, and
    Department of Physics, University of Colorado, Boulder, Colorado 80309, USA}

    \date{\today}%
    
    \begin{abstract}
    
    We present techniques to model and design membrane phononic crystals with low-mass defects, optimized for force sensing.  Further, we identify the importance of the phononic crystal mass contrast as it pertains to the size of acoustic bandgaps and to the dissipation properties of defect modes. In particular, we quantify the tradeoff between high mass contrast phononic crystals with their associated robust acoustic isolation, and a reduction of soft clamping of the defect mode.  We fabricate a set of phononic crystals with a variety of defect geometries out of high stress stoichiometric silicon nitride membranes, and measured at both room temperature and 4 K in order to characterize the dissipative pathways across a variety of geometries. Analysis of these devices highlights a number of design principles integral to the implementation of low-mass, low-dissipation mechanical modes into optomechanical systems. 
    
    \end{abstract}
    
    \maketitle
    
    \section{Introduction}

Many nanoscale sensing and transduction protocols are based on detection of the minuscule forces between a sample and a highly sensitive probe. Examples include microscopy methods such as atomic force microscopy (AFM) and magnetic resonance force microscopy (MRFM)~\cite{rugar2004single,poggio2010force-detected}. Additionally, an increasing number of quantum protocols make use of a mechanical intermediary to transduce between two types of otherwise non-interacting quantum systems~\cite{safavi-naeini2010proposal,higginbotham2018harnessing,rabl2010quantum}.  Readout of the mechanical motion can be achieved through a range of optical, electrical, or magnetic means~\cite{aspelmeyer2014cavity,ekinci2005nanoelectromechanical}. Fundamentally, all these techniques are often limited by environmental, i.e.~thermal or Brownian, noise coupled to the mechanical resonator.  This noise leaks into the mechanical mode at the mechanical damping rate as dictated by the fluctuation-dissipation theorem, defining a force sensitivity noise floor.

The factors that define mechanical damping to the environment have been a long-standing question.  Generally, loss mechanisms can be divided into two categories:  Acoustic radiation into the external structure~\cite{wilson2008intrinsic}, and internal loss (dissipation) of acoustic motion to heat through bending -- either at the clamping point, or within the structure itself. Historically, control of materials has been a main driver at reducing loss. Recently, progress is increasingly propelled by engineering of geometry and structural parameters to mitigate both mechanisms.  Silicon nitride tensioned thin films are a platform that offer multiple control parameters~\cite{verbridge2008megahertz,zwickl2008high,unterreithmeier2010damping,yu2012control,yuan2015silicon,norte2016mechanical,reinhardt2016ultralow-noise,barasheed2016optically,tsaturyan2017ultracoherent,ghadimi2018elastic}. Tension of the film results in dissipation dilution, which maintains low dissipation while increasing the oscillator energy.  Acoustic isolation in the form of bandgap engineering of the surrounding substrate can control acoustic radiation~\cite{alegre2010quasi-two-dimensional,yu2014phononic,tsaturyan2014demonstration}, and in recent realizations nanopatterning of the tensioned SiN film itself can provide both (1) a phononic bandgap and (2) the phenomenon called soft clamping, which describes the gradual decay of the mechanical mode into the phononic crystal (PnC) structure reducing bending loss~\cite{ tsaturyan2017ultracoherent,ghadimi2018elastic}.

In this work we explore PnCs with varying mass contrast and associated defect designs in two-dimensional periodic structures. We observe that increasing mass contrast of the PnC widens the bandgap in the frequency domain [Fig.~1 (a,b)]; at the outset this improves isolation of the mechanical motion and avoids contamination of the spectrum due to thermal noise of the integrated motion of other modes. Further, designs with high mass ratio are amenable to analytic analysis, as has been carried out recently in the context of 1D PnC strings~\cite{ghadimi2018elastic}. This design methodology simplifies the design of a range of defects, in particular, smaller effective mass defects and hence better force sensitivity. Small-mass defects can also be incorporated into low-contrast PnCs, through insight from FEM simulations, and we study a range of defect designs and measurements in low-contrast PnCs~\cite{tsaturyan2017ultracoherent}.

When increasing mass contrast of PnCs mode localization is enhanced, but a consequence is a reduction in soft clamping, which increases susceptibility to internal (bending) loss. Here we consider this decrease in soft clamping quantitatively by examining the bending of multiple combinations of intertwined PnC and defect designs. We note that the choice of tradeoff between localization and soft-clamping will depend upon the application considered; when operating at room temperature soft-clamping is critical, but when at deeply cryogenic temperatures~\cite{yuan2015silicon} or in crystalline materials~\cite{cole2014tensile,buckle2018stress} the associated small internal loss rates may allow for sufficiently low dissipation even when the mode is strongly localized.

For designs in this work, we specifically have in mind applications where high force sensitivity is desired for MHz-scale membrane resonators simultaneously coupled to an optical mode and a spin or electrical degree of freedom~\cite{scozzaro2016magnetic,fischer2019spin,andrews2014bidirectional}. For these functionalization purposes, we considered a few design constraints. This include designing defects large enough to couple to Fabry-Perot cavities~\cite{purdy2012cavity}, fabricating wide bandgaps that can incorporate a wide variety of defects robust to perturbations due to deposition of metallic or magnetic components, as well as limiting device size to avoid instabilities resulting from low frequency mechanical modes. 

We begin the article with an analytic and numerical analysis of design considerations for optimizing both bending and radiative loss.  In particular, we correlate the PnC mass contrast to bandgap width, phononic isolation, and variation in mechanical quality ($Q$) due to corresponding changes in soft clamping. In this way, we link the geometrical design considerations needed to achieve soft clamping to a relatively narrow bandgap. We then present measurements of a set of five devices with varying bandgap and defects; we study $Q$ at both room temperature and 4 K where internal loss is the dominant contributor, such that we can compare the experimental results to loss prediction from bending. As a primary metric, we examine the mechanical force sensitivity of all devices, developing design principles for combinations of both low mass and low loss mechanical defects. We also identify structures in which nuances of defect design has deleterious consequences, such as considerable defect bending, or large effective mass. In this study we utilize 100 nm thick SiN PnCs and demonstrate force sensitivities down to 12 $\mathrm{aN}/\sqrt{\mathrm{Hz}}$ at temperatures of 4 K and a frequency of 1.6 MHz. We expect that thinner films and colder temperatures will further enable optomechanics that combines high frequency membranes with excellent force sensitivity~\cite{nichol2012nanomechanical,yuan2015silicon,tsaturyan2017ultracoherent,reinhardt2016ultralow-noise,norte2016mechanical}.

    \begin{figure}[t]
        \centering
        \includegraphics[width=80mm]{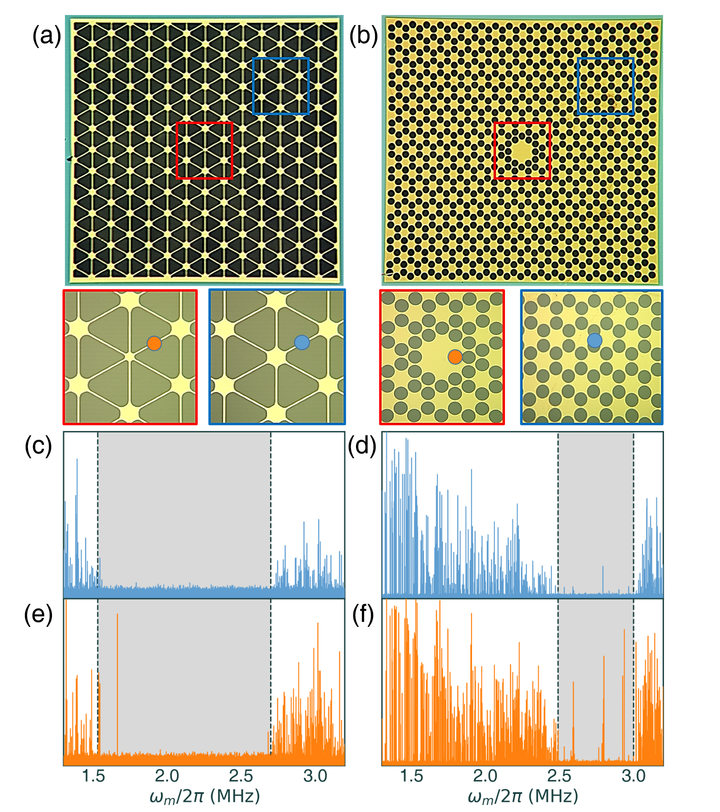}
        \caption{(a,b) Optical microscope image of high and low mass contrast PnC devices containing a defect respectively. The displacement of the devices measured both in the crystal bulk (blue) and at the defect (orange).  (c,e) Thermal noise spectra of the high contrast device on a logarithmic scale. The defect mode shown in (e) is the fundamental symmetric mode. (d,f) Mechanical spectra for the low contrast device on a logarithmic scale. The device was driven with sufficient white noise in order to observe all expected defect modes. The spectrum shown in (f) contains 5 separate defect modes within the bandgap, including the second symmetric mode examined in this study. Defect mode frequencies in both (e) and (f) agree with predictions from 2D FEM simulations. Mechanical modes that appear inside the bandgap in (d) have quality factors less than 100, and therefore are most likely hybridized modes between the membrane and the silicon chip or mounting assembly.}
        \label{fig:pnc}
    \end{figure}
    
    \section{Concept and PnC Modeling}

    Here we present the design and analysis of two dimensional PnC structures and their defects. The defect is designed such that it has one or more mechanical mode within the bandgap of the surrounding PnC. Such a design supports a spatially localized mechanical mode within the larger patterned membrane structure. The design of all devices in this work is guided by an optimization of the force sensitivity~\cite{saulson1990thermal}: 

    \begin{equation}
        S_{FF} = \frac{4 k_{\mathrm{B}} T m \omega_{\mathrm{m}}}{Q} =  \frac{2 \hbar k_{\mathrm{B}} T}{Q x_{\mathrm{zp}}^{2}} 
    \end{equation}
    Here, $x_\mathrm{zp}$ is the resonator's fluctuation at zero absolute temperature, $\omega_{\mathrm{m}}$ is the angular frequency of the mechanical mode, $T$ is temperature, $k_B$ is the Boltzmann constant, and $m$ is the effective mass of the mode. We emphasize that the appearance of $x_{\mathrm{zp}}$ is not an indication of quantum effects, but rather a convenient parameterization as $x_{\mathrm{zp}}$ is a fundamental parameter in both the force sensitivity and the optomechanical coupling~\cite{aspelmeyer2014cavity}. We see that force sensitivity can be enhanced by maximizing both $x_{\mathrm{zp}}$ and $Q$. In pursuit of the latter, we note that dissipation can be expressed as a sum of two contributions
    
    \begin{equation}
    \frac{1}{Q} = \frac{1}{Q_{\mathrm{bend}}}+\frac{1}{Q_{\mathrm{rad}}}
    \end{equation}
    where $Q_{\mathrm{bend}}$ arises from the internal, or bending losses of the mechanical mode, while $Q_{\mathrm{rad}}$ encapsulates acoustic radiation that is lost to the substrate and wider environment. A SiN PnC can enhance both $Q_{\mathrm{rad}}$ and $Q_{\mathrm{bend}}$ via acoustic isolation and soft clamping, respectively.
    
    \subsection{1D Model for Phononic Crystals}
    
\begin{figure}[h]
    
    \includegraphics[width=80mm]{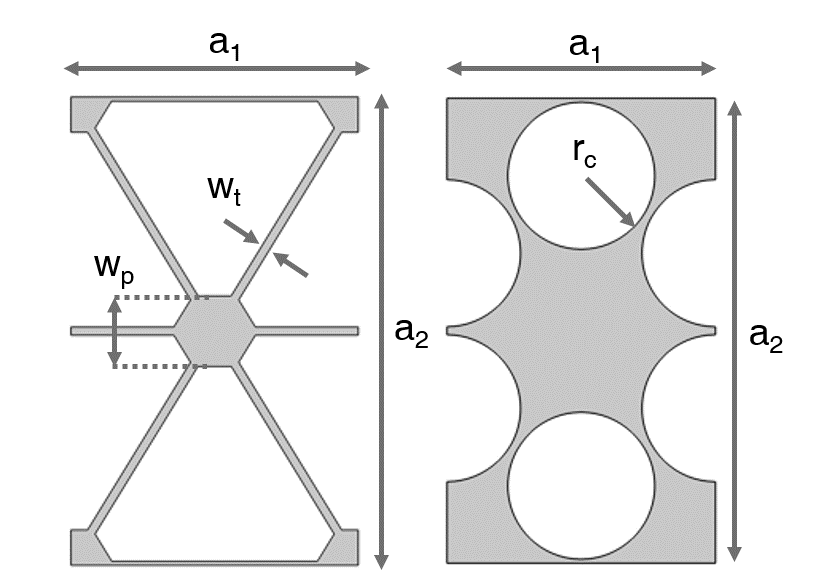}
    \centering
    \def\arraystretch{1.2}
    \setlength{\tabcolsep}{2.5pt}
    \begin{tabular}{llllllll} 

     \hline
     Device& $a_1$ & $a_2$ & $n_x$ & $n_y$ & $\mathrm{w}_p$ & $\mathrm{w}_t$ & $r_c$\\ 
    \hline
     A [Fig.~1(a)] & \SI{100}{\micro \meter} & \SI{172}{\micro \meter} & 12 & 7 & \SI{30}{\micro \meter} & \SI{2.5}{\micro \meter} & --\\
     B [Fig.~1(b)]  &\SI{87}{\micro \meter} & \SI{151}{\micro \meter} & 19 & 11 & --& -- & \SI{23}{\micro \meter} \\ 
     C &\SI{111}{\micro \meter} & \SI{192}{\micro \meter} & 9 & 5 & --& -- & \SI{31}{\micro \meter} \\
     D & \SI{67}{\micro \meter} & \SI{116}{\micro \meter} & 15 & 9 & --& -- & \SI{18}{\micro \meter} \\ 
     E &\SI{100}{\micro \meter} & \SI{172}{\micro \meter} & 12 & 7 & \SI{30}{\micro \meter} & \SI{2.5}{\micro \meter} & -- 
     \\
     \hline
    \end{tabular}
    \caption{Critical dimensions for the unit cells of all fabricated devices. Dimensions in table are defined above. $n_x$, $n_y$ are the unit cell number of the PnC along the X, Y axes. Table entries are omitted where they do not apply. Fabricated devices for A and C included fillets of radii \SI{2.5}{\micro \meter} at the sharp corners of the pad.}
    \label{fig:pnc_design}
    \end{figure}

  In this work we are ultimately interested in the design and fabrication of 2D structures. However, the main design methodologies and concepts can be elucidated by considering a 1D analysis where we convert 2D unit cell (Fig.~\ref{fig:pnc_design}) to a 1D unit cell. In this work, we separate each unit cell into regions of thin tethers and wide pads as depicted in Fig.~\ref{fig:pnc}a. In this geometry, the pads (tethers) will have slower (faster) wave velocities than the unpatterned membrane. Periodic modulation of the wave velocity results in gaps in the acoustic spectrum. This procedure, in principle, integrates over the mass of the transverse dimension, and recalculates the appropriate wave velocities. Details are outlined in Appendix A.

    We present analysis of both finite and infinite PnC structures. A complete description of finite structures involves solving the 1D Euler-Bernoulli equation for the out of plane displacement $u(x,t)$, which describes the bending of the PnC beam. Such an analysis captures the behavior of the mode shape near clamping points of the finite structure~\cite{fedorov2019generalized}
    
    \begin{equation}
        \dfrac{d^2}{dx^2}\big[I(x) E \dfrac{d^2 u(x)}{dx^2}\big] - \mathcal{T} \dfrac{d^2 u(x)}{dx^2} - \rho_{\mathrm{1D}} \dfrac{d^2 u(x)}{dt^2} = 0
    \label{eq:eulerBern}
    \end{equation}
    where $I(x)$ is the geometric moment of inertia, $E$ is the Young's Modulus, $\mathcal{T}$ is the tension, and $\rho_\mathrm{1D}$ is the linear mass density. It can be seen that for high aspect ratio devices, the 4th order term is negligible away from the clamps ~\cite{fedorov2019generalized}. 
    
    For infinite structures, we can assume periodic boundary conditions. Thus, any expected mode shape should have minimal contribution from the bending term in the Euler-Bernoulli equation. Under these conditions the Euler-Bernoulli equation reduces to the 1D wave equation, and thus the motion of the string is completely parameterized by its spatially dependent wave velocity:
    \begin{equation}
        v_{p,t} = \sqrt{\frac{\mathcal{T}}{\rho h \tilde{\mathrm{w}}_{p,t}}}
        \label{eq:velocity}
    \end{equation}
    where $\rho$ is the bulk density of the membrane, $h$ is the membrane thickness, and $\tilde{\mathrm{w}}_{\mathrm{p}}$ ($\tilde{\mathrm{w}}_{\mathrm{t}}$) is the converted pad (tether) width. Definitions of the converted pad and tether widths appear in Appendix A.
    
    We outline the solution for infinite structures in Appendix B. As is typical in these types of problems, the band structure is parameterized by a transcendental equation:
    
    \begin{equation}
        \arccos{S} = a k
    \label{eq:transcendental}
    \end{equation}
    \begin{equation}
    \begin{split}
        S = \frac{(V+1)^2}{4V}\cos{\big[\omega (t_p + t_t)\big]}\\ - \frac{(V-1)^2}{4V}\cos{\big[\omega (t_p-t_t)\big]}
    \end{split}
    \label{eq:S}
    \end{equation}
    where $t_{\mathrm{p,t}}=l_{\mathrm{p,t}}/v_{\mathrm{p,t}}$  are the transit times of acoustic waves through the pad and tether respectively, $a$ is the unit cell length, $k$ is the Bloch wavenumber, and $V \equiv v_t/v_p$ is what we define as the contrast. As can be seen in Eq.~\ref{eq:velocity}, the contrast $V$ is geometrically defined by the mass contrast, or the relative converted width of the pads and tethers. Traveling wave solutions are prohibited when $|S| > 1$; such a condition defines the range of $\omega$ that determines all bandgaps.

    One goal of PnC design is to maximize the bandgap size in the frequency spectrum. For simplicity, we investigate the case where we wish to maximize the width of the first bandgap. Inspection of Eq.~\ref{eq:S} reveals that this occurs when the second term is identically 0. In this case, $t_p = t_t$, and $|S|$ reaches its maximal value $S_{\mathrm{max}}$ when $\omega = \omega_{\pi} \equiv \pi / 2 t_t$:
    \begin{equation}
        S_{\mathrm{max}} = \frac{(V+V^{\minus 1}+2)}{4}
        \label{eq:Smax}
    \end{equation}
    With this in mind, we can derive an expression for the normalized bandgap width:
    \begin{equation}
        \Delta = \frac{\Delta \omega}{\omega_{\pi}} =  \ \frac{2}{\pi}\arccos{\bigg( \minus \frac{1-6V+V^2}{1+2V+V^2}\bigg)}
        \label{eq:Delta}
    \end{equation}

\begin{figure}[t]
        \centering
        \includegraphics[width=80mm]{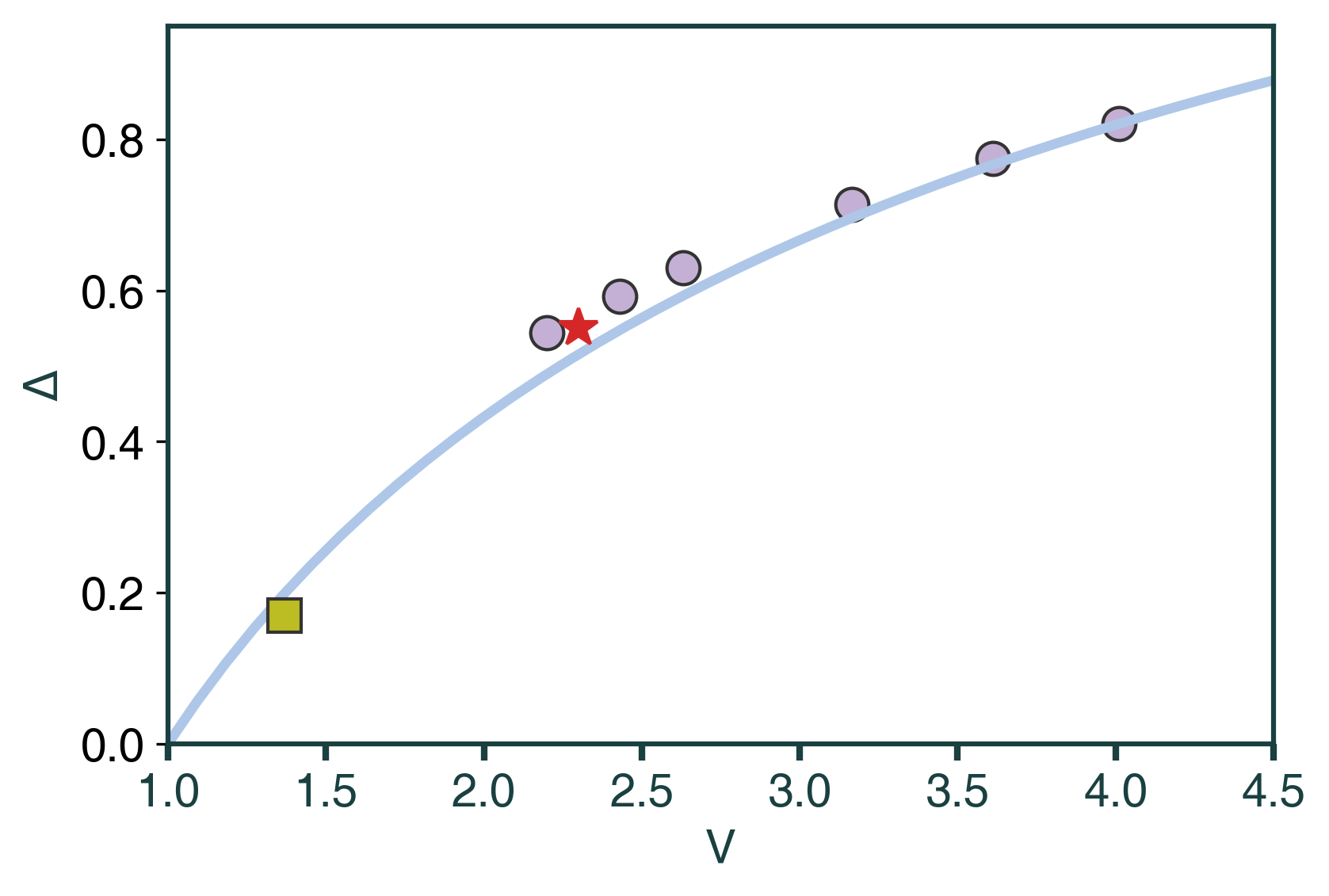}
        \caption{Band gap widths as a function of mass contrast as determined from the 1D model. Comparison between different models of the bandgap to mid-gap ratio ($\Delta$) versus PnC contrast ($V$). Analytic prediction of the 1D model (Eq.~\ref{eq:Delta}) (solid blue line), finite element simulations of equivalent 2D structures (blue circle markers), experimental results of device A (red star marker) and device B (yellow square).}
        \label{fig:concept}
    \end{figure}

    A strong agreement between the 1D model prediction of Eq.~\ref{eq:Delta} (blue line) and FEM simulations of equivalent 2D unit cell (points) across a range of contrasts can be seen in Fig.~\ref{fig:concept}. The main source of error in the 2D to 1D conversion arises from the inability of the 1D model to correctly account for the transition between the low mass tethers and high mass pads of the 2D PnC. In the high mass ratio limit, these regions account for a smaller fraction of the geometry, and thus the 1D model better captures the resulting band structure for large $V$. Figure~\ref{fig:concept} also displays results for fabricated devices. It is readily seen that the device illustrated in Fig~\ref{fig:pnc}(a) (red star) has a higher mass contrast compared to the device in Fig~\ref{fig:pnc}(b) (yellow square). This difference in mass contrast is also reflected in the corresponding bandgap width of both devices.
    
    A technical point arises in the fabrication of devices in the limit of a very large bandgap. As said before, $V$ is ultimately is determined by the relative sizes of the pads and tethers, with wide pads and narrow tethers giving rise to large values of $V$. Therefore, the magnitude of $V$ is ultimately determined bt the minimum tether width that can be fabricated and the desired bandgap position. As an example, a PnC with $V = 4$ would have pad widths of \SI{40}{\micro \meter} assuming a $\mathrm{w}_{\mathrm{t}} = $ \SI{1}{\micro \meter}. The principle of equal transit times then sets the unit cell length to be \SI{220}{\micro \meter}, which in turn sets $\omega_{\pi} = 2 \pi \times$\SI{1}{\mega \hertz}. In general, higher contrast (higher $V$) phononic crystals will have lower bandgap center frequencies, because the larger mass ratio induces lower average wave velocities. 

\subsection{Effects of Contrast on Soft Clamping}  
    
    The main advantages afforded by SiN PnCs include both soft clamping and phononic isolation of defect modes. We assert that $V$ (and thus also $\Delta$) is a geometrically defined indicator of the extent to which a PnC achieves these phenomena. To understand the effects of soft clamping on dissipation, we first examine the origin of bending loss for a given mechanical mode~\cite{unterreithmeier2010damping,schmid2011damping,yu2012control}:

    \begin{equation}
        Q_{\mathrm{bend}} = Q_{\mathrm{int}}(h) \frac{24 (1-\nu ^2)}{E h^3} L_{\mathrm{s}}^{\minus 1} = \frac{24 (1-\nu^2)}{E_2(h) h^3}L_{s}^{\minus 1}
        \label{eq:bending}
    \end{equation}
    \begin{equation}
        L_{\mathrm{s}} = \int{}^{} \kappa(x,y) dx dy
    \end{equation}
    
    \begin{equation}
        \kappa =\frac{1}{U}(\partial_x^2 u(x,y) + \partial_y^2 u(x,y))^2
    \end{equation}
    where $Q_{\mathrm{int}}(h)$ is the intrinsic $Q$ of silicon nitride, $E$ is the Youngs' mondulus, $h$ is the nitride thickness, $\nu$ is the Poisson ratio, $U$ is the kinetic energy of the mode and $\mathrm{u(x,y)}$ is the mode shape. $E_{2}(h)$ is the imaginary part of the Youngs' modulus, defined to be $E_{2} = E / Q_{\mathrm{int}}$. Here we call attention to the dependence of $Q_{\mathrm{int}}$, which has been shown to have a scaling $Q_{\mathrm{int}} \propto h$ at membrane thicknesses around 100 nm \cite{Villanueva2014evidence}. We will call $L_s$ the loss factor, which quantifies how much bending losses a given mode has. We also define $\kappa$ as the loss density, which describes the spatial distribution of lossy motion of a mode. Therefore bending loss in membrane devices is most pronounced at clamping points where the mode shape curves strongly to adhere to the restrictive clamping boundary conditions. 
    
    \begin{figure}[h]
    \centering
        \includegraphics[width=85mm]{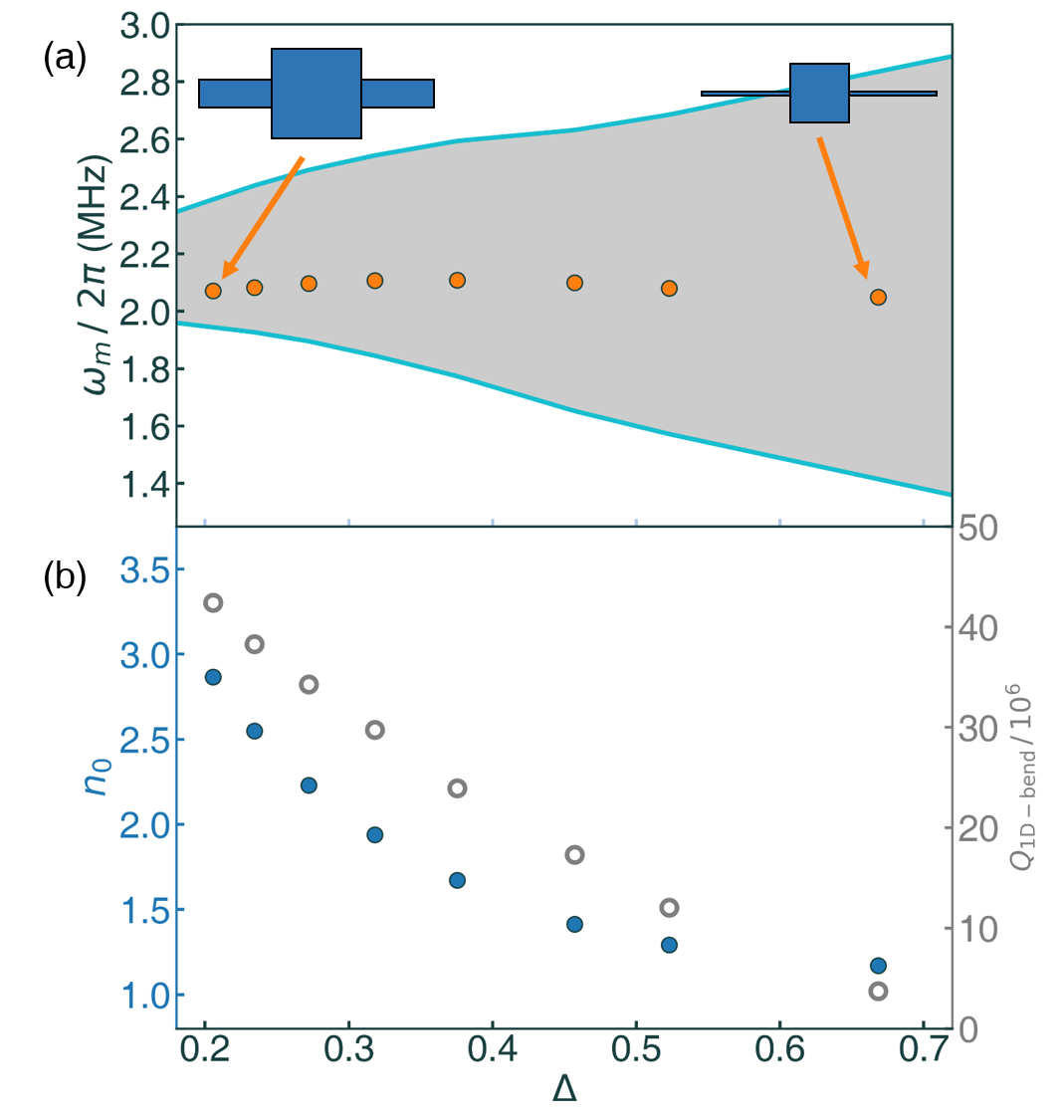}
        \caption{Effects of contrast for 1D PnCs with a constant defect geometry. (a) Orange points are the fundamental defect mode frequencies across all simulated structures. The shaded region denotes the frequency range of the simulated bandgaps. The inset in (a) are schematics of the simulated 1D unit cell at the lower extreme (left) and upper extreme (right) of the bandgap width. (b) Normalized mode amplitude decay length (blue) and quality factors (grey) for 1D PnC strings with a constant defect. We define $n_0=l_0 / a$, where $l_0$ is the exponential decay length. At large contrasts $n_0$ approaches $1$ because the defect mode does not begin decay until the onset of the PnC.}
        \label{fig:softclamp}
    \end{figure}
    
    Soft clamping and phononic isolation occur when the mechanical defect mode has a frequency within the acoustic bandgap of the PnC. Within the bandgap, $k$ become complex. The magnitude of $\operatorname{Im}(k)$ at the defect mode frequency determines the decay length $l_0 = 1/\operatorname{Im}(k)$ of the defect mode into the PnC.

    To directly illustrate how PnC design affects the soft clamping of defect modes, we perform a series of 1D simulations where a single defect is placed in a series of PnCs with ascending contrast. From the calculated defect mode shapes, one can predict a value for $Q_{\mathrm{bend}}$ assuming that $Q_{\mathrm{int}}$ has a value of 6600~\cite{Villanueva2014evidence}. In this 1D simulation, the defect mode's frequency has a nearly constant position relative to the bandgap center [Fig.~\ref{fig:softclamp}(a)]. Two effects as a function of increasing $\Delta$ are apparent. First, the normalized mechanical decay length $n_0 = l_0 / a$ decreases as a function of $\Delta$ [Fig.~\ref{fig:softclamp}(b)], as predicted from our analytic analysis in the 1D model. Second, a decrease of $Q_{\mathrm{bend}}$ is apparent.  This is indicative of the bending resulting from the strong decay into the PnC.  
    
    \subsection{Effect of Contrast on Phononic Isolation}
    
\begin{figure}[b]
        \centering
        \includegraphics[width=80mm]{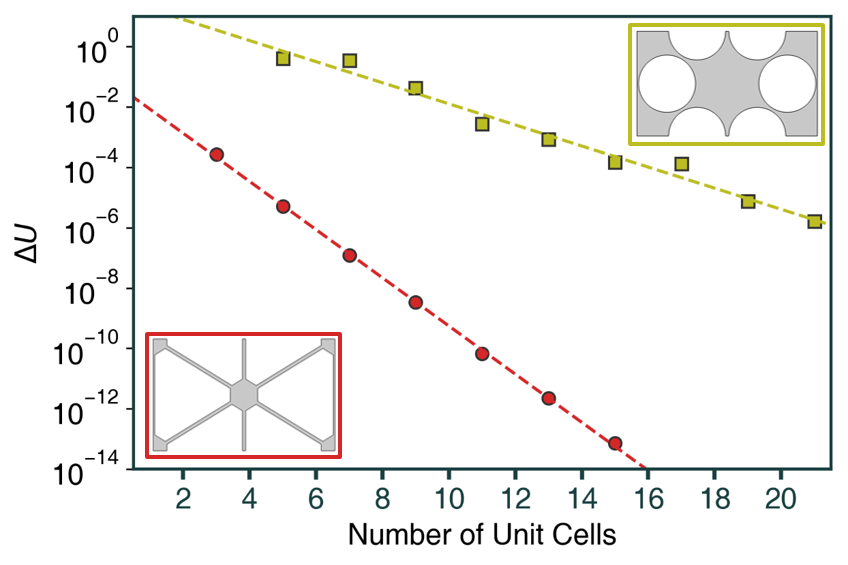}
        \caption{Kinetic energy in the PnC frame, normalized to the total energy, in both a high contrast (red points, $V=2.3$) and low contrast (yellow points, $V=1.3$) PnCs as calculated from FEM simulations. Insets are the unit cells used in the simulation. Dashed lines are fits assuming exponential energy decay for the defect mode. The decay length for the high contrast unit cell (red fit) is 0.54 unit cells. The decay length from the low contrast unit cell (yellow fit) is 1.24 unit cells.}
        \label{fig:isolation}
    \end{figure}

    The numerical analysis above showed that high contrast PnCs exhibit short defect mode decay lengths, and we also see numerically that this short decay length provides enhanced suppression of radiative loss. In this analysis, we investigate the concept of phononic isolation in a two dimensional structure by FEM simulation using Comsol. For details, see Appendix D. Here, we calculate the ratio of the tensile energy stored in the entire structure to the energy stored in a \SI{10}{\micro \meter} overhang around the PnC perimeter. We define this ratio $\Delta U$. As seen in Fig.~\ref{fig:isolation}, the higher contrast PnC has a faster energy decay compared to the low contrast crystal. We note that in the 1D case, $S_{\mathrm{max}}$ is reached when the defect mode frequency is at the center of the bandgap. However, even if this condition is not obtained, one can still achieve robust isolation in the high contrast limit.

    \subsection{Beyond the 1D Model:  Incorporating Defects in 2D} 
    
    Up to this point, we focused on PnC characteristics, and have shown that our 1D model reproduces the the basic behavior the equivalent 2D crystals. However, the same cannot be done for designing defects in two dimensions, where we generally find that the 1D results have at most a qualitative relationship to a 2D defect. This is mostly due to the complicated stress redistribution that occurs in intricate 2D structures around the defect (Fig.~\ref{fig:summary}), which a 1D model cannot fully capture. This internal structure of the defect leads us to further divide the loss pathways in our 2D devices into losses that can encapsulated in 1D simulations ($Q_{\mathrm{1D-bend}}$) and losses that are inherently 2D ($Q_{\mathrm{2D-bend}}$). In this next section, for example, we designed 5 different devices with a various combinations of defect and PnC designs (Fig.~\ref{fig:summary}). FEM simulations of the defect mode shape help illustrate the difference between the aforementioned separation between ``1D loss" and ``2D loss". For example in the context of high-contrast PnCs, a linecut of the defect mode of device A resembles closely a string mode emanating radially from the PnC center. Device E has a large defect pad, and therefore there is considerable internal motion of the pad. This internal motion gives rise to a larger participation ratio of the ``2D loss" as compared to device A. Therefore, one expects that 1D analysis is much more applicable to device A than to device E. 
    
    The stress redistribution of the 2D structure also strongly affects the position of the defect mode frequency within the bandgap, leading to  discrepancies between the 1D and 2D simulations.  While the 1D model allows positioning the defect modes directly in the bandgap center [Fig.~\ref{fig:softclamp}(a)], the equivalent mode has a lower frequency in both 2D FEM simulations and the experimental results [Fig.~\ref{fig:pnc}(e)].

    \section{Experimental Results}

    \begin{figure*}[t!]
        \includegraphics[width=160mm]{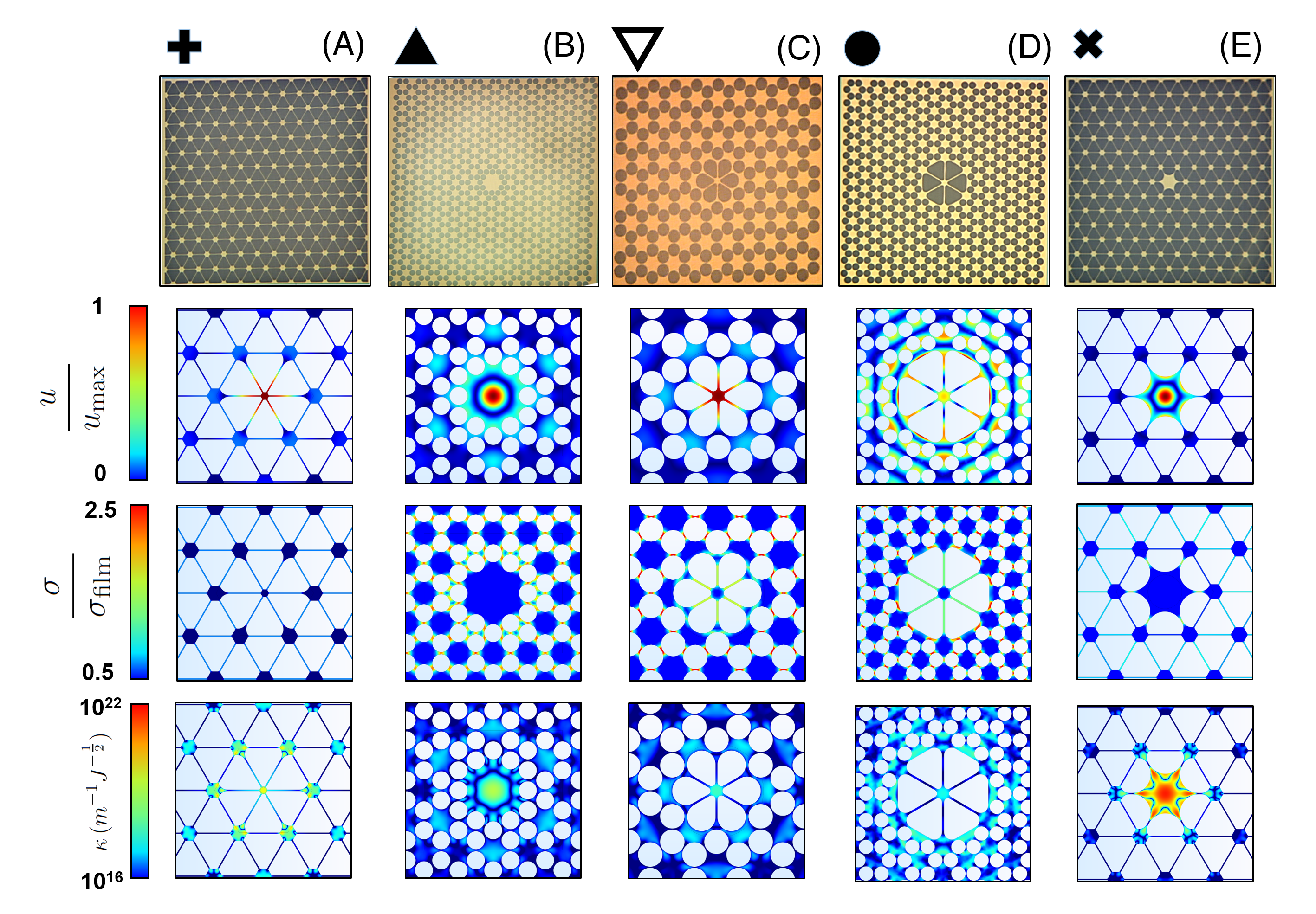}
        \caption{Compilation of all devices presented in this work. In the first row are optical microscope images of suspended phononic structures. Above are device labels that correspond to device properties plotted in Fig.~\ref{fig:Sfplot}. The second row presents FEM simulations of symmetric defect mode shapes. The third row shows FEM simulations of the static stress distribution normalized to the film stress. The fourth row displays the normalized bending loss density.}
        \label{fig:summary}
    \end{figure*}
    
    \begin{figure}[h]
        \centering
        \includegraphics[width=85mm]{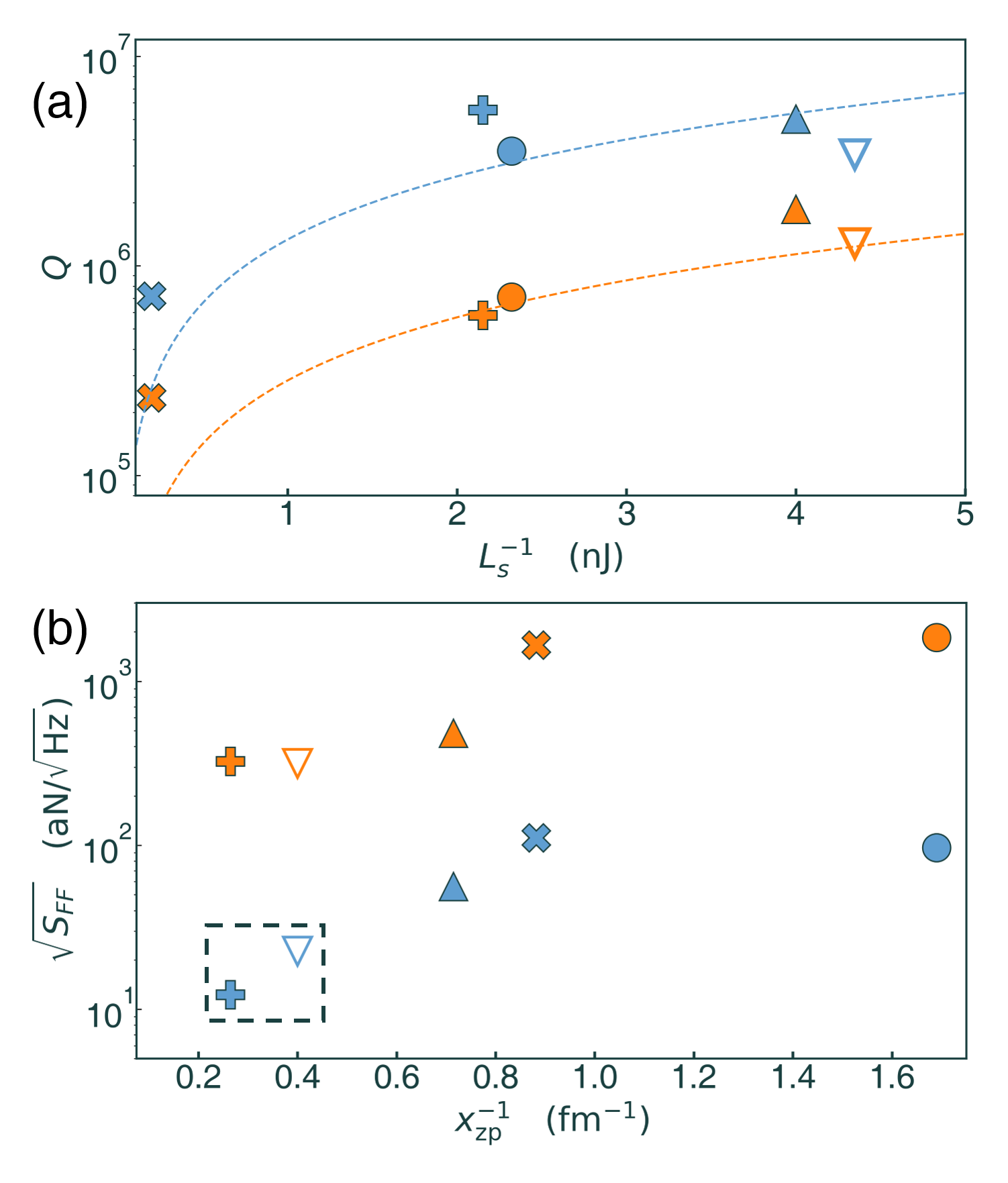}
        \caption{(a) Quality factor dependence on the loss factor $L_s$ for all fabricated devices, measured at room temperature (orange), and at 4 K (blue). The symbols correspond to the key presented in Fig.~\ref{fig:summary}. Dashed lines represent calculated $Q$-values from Eq.~\ref{eq:bending}, with $E_2 =\,$ \SI{80}{\mega \pascal} for the room temperature data and $E_2 =\,$ \SI{17}{\mega \pascal} for 4 K. (b) Calculated force sensitivity at room temperature (orange) and at cryogenic conditions (blue). The box is to emphasize the similarity in defect design and performance between devices A and C.}
        \label{fig:Sfplot}
    \end{figure}
    
    To test examples of the above ideas, we fabricated and characterized multiple devices (Fig.~\ref{fig:summary}) utilizing the two different PnC designs illustrated in Fig.~\ref{fig:pnc} and defined in Fig.~\ref{fig:pnc_design}, and multiple defect designs. The chosen devices illustrate the complex interplay of bending loss (both in the PnC and defect) and effective mass $m$ (characterized by $\xzp$) of the defect mode. The phononic devices were patterned into a 100 nm thick LPCVD silicon nitride layer on a silicon substrate, and suspended using a KOH wet etch. All Si substrates were 375 $\mu$m thick 5$\times$5 mm$^2$ in the transverse dimensions. Full fabrication information appears in Appendix C.
    
    The mechanical mode spectrum of the devices was characterized at both room temperature and 4 K by affixing the device to a mirror mounted on a piezoelectric transducer. The spectrum was imprinted on the amplitude fluctuations of light reflected from the etalon formed by the mirror and the membrane device. For all measurements, around 5 mW of 1064 nm laser power was incident on the sample. An absolute calibration of the mechanical displacement of our devices can be performed by assuming that the observed mechanical amplitude is entirely due to the expected thermal signal. However, many of our spectra were taken via white noise driving of the entire PnC chip via the piezo electric transducer. Therefore, all mechanical amplitudes are displayed relative to the shot noise limited noise floor the detection chain. The room temperature measurements were performed at a pressure less than $\mathrm{10^{-6} \, mBar}$. 4 K measurements were performed in a closed cycle cryostat with free space optical access.
    
    To clearly understand the mode spectrum we characterize the motion in multiple regions of the PnC. Outlined in Fig.~\ref{fig:pnc}, the different colored spectra were obtained by probing near the defect (orange point) and in the crystal bulk (blue point). Modes that only appear when probing near the defect are confined by the PnC and considered to be defect modes [Fig~\ref{fig:pnc}(e,f)]. When probing in the crystal bulk [Fig~\ref{fig:pnc}(c,e)], one sees a wide range of frequencies with no discernible mechanical resonance (grey regions in Fig.~\ref{fig:pnc}(c-f). This frequency range is found to be broader for the high contrast PnC pictured in Fig.~\ref{fig:pnc}(a), and is accordance with the results of FEM simulations. We note that in order to adequately resolve the upper bandgap edge, we chose to measure on a tether as it accommodates more high frequency motion above the bandgap than in the pads of the PnC. Occasionally, spurious modes were found inside the bandgap (Fig.~\ref{fig:pnc}). These modes had quality factors on the order of 100, and therefore are inconsistent with pure silicon nitride membrane modes. This is consistent with the fact that these are hybridized modes between the membrane and its substrate or the chip-mirror assembly~\cite{yu2014phononic}.
    
    Quality factors were determined via ringdown measurements of the phononic structures driven by the piezoelectric transducer. Figure \ref{fig:Sfplot}(a) displays the quality factors for the symmetric modes of all devices at both room temperature and 4 K. Loss factors were calculated with Eq.~\ref{eq:bending} by FEM simulations of each device. The fits are based on the assumption that the dissipation is bending loss limited $Q_{\mathrm{rad}} \ll Q_{\mathrm{bend}}$. For the room temperature data, the fit value of $E_2 = \,$\SI{80}{\mega \pascal} combined with the assumption of $E = \,$\SI{250}{\giga \pascal} infers $Q_{\mathrm{int}} = 3125$. This has agreement tabulated values of $Q_{\mathrm{int}}$ found in the literature \cite{Villanueva2014evidence}. The fit value of $E_2 = \,$\SI{17}{\mega \pascal} at 4 K infers $Q_{\mathrm{int}} = 14700$. We note that scatter of points, particularly at room temperature, indicates that this assumption may not hold for all devices. Notably, device C which has a rather large value of $\Delta U$, which may be indicative of the existence of measurable external loss for this device.

    In Fig.~\ref{fig:Sfplot}(b) we see the force sensing performance across all devices at both room temperature and 4 K. As discussed previously, we use force sensing performance of these high-frequency resonators as a comparison metric.  First, let us comparatively examine devices A and B. Device A displayed among the best performance among all devices, with a force sensitivity of 12 $\mathrm{aN}/\sqrt{\mathrm{Hz}}$ at 1.6 MHz at 4 K. Both devices A and B contain a well confined defect mode, leading to a relatively large $x_{\mathrm{zp}} > 1 \, \mathrm{fm}$ at MHz frequencies. This confinement is accomplished simultaneously with little dissipation. For device A (Fig.~\ref{fig:summary}) the bending loss of the fundamental trampoline mode is most pronounced in the pads surrounding the defect itself. This type of bending profile is similar to that seen in SiN trampolines~\cite{reinhardt2016ultralow-noise,norte2016mechanical}, while relaxing some of the clamping effects at the edge of the defect mode. Device B maintains low dissipation via the soft clamping of its surrounding low contrast PnC. Although both devices demonstrate relatively high force sensitivity, they differ considerably in design, and therefore include clear design trade-offs. Device A has low effective mass due to its low physical mass. Additionally, the low effective mass of Device A can also be attributed to its first order symmetric mode, which is generally lower in mass since it has a node of motion at the defect boundary. However, its frequency is close to the bandgap edge (Fig.~\ref{fig:pnc}). This frequency is most strongly influenced by the length of the defect tethers~\cite{norte2016mechanical}, and thus frequency cannot be easily pushed higher.  Device B has a second order symmetric defect mode. In general, these second order modes are easier to place in the bandgap center, but are higher mass because the edge of the defect coincides with an antinode of motion.
    
    To compare the effects of the PnC on the defect mode, we can compare devices A and C, which both have defects with trampoline geometries. Both devices also support a first asymmetric defect mode with frequencies between 1.5 and 2 MHz. We find overall that these devices exhibit a similar force sensitivity both experimentally [Fig.~\ref{fig:Sfplot}(b)] and in FEM simulations.  There are however slight differences in how these devices achieve their sensitivity.  Finite element simulations predict that device C will have more soft clamping and hence higher $Q$ than device A (although in experiment device A has higher Q, perhaps due to its larger number of unit cells). However, device A has lower mass, and hence larger $\xzp$, than device C. Note, even when the defect pad width of device C is set to be identical to that of A, the increase in pad width is associated with an increase in resonant frequency and the same conclusions hold.
    
    The remaining devices (D and E) provide examples of how nuances in device design can cause adverse effects. For instance, device D exhibits a relatively low $\xzp$ [Fig.~\ref{fig:Sfplot}(b)]. Here, the defect accommodates the second symmetric mode, causing the PnC to carry a large portion of the defect motion (Fig.~\ref{fig:summary}). This differs from the fundamental trampoline-like mode shape in that the second symmetric mode has an anti-node of motion at the PnC boundary. Thus the motion of the defect strongly drives the motion of the surrounding low contrast crystal adding considerable mass to the mode. This effect is not seen in device C (same PnC, different defect), because this defect was designed to accommodate the fundamental mode within the bandgap. Finally,  device E is composed of a large defect pad intended to isolate the second symmetric mode through the PnC. which leads to a large amount of defect loss. The large pad is largely stress released, and thus the edges of the pad are allowed to freely bend (Fig.~\ref{fig:summary}). We can compare this to device A which has a large released pad, but it is surrounded by a highly stressed boundary. The relieved stress regions accommodate extreme bending which leads to substantial internal loss, thus limiting the $Q$.  This effect has ramifications for future PnC design, where a large pad is often desired for efficient coupling between an optical mode and the mechanical motion. 
    \begin{figure}[h]
    \def\arraystretch{1.2}
    \setlength{\tabcolsep}{2.5pt}
    \begin{tabular}{ lllllll }
     \hline
     Device& $\mathrm{\omega_m} / 2 \pi$ & $\mathrm{x_{\mathrm{zp}}}$& $\mathrm{\Delta}$ & $\mathrm{\Delta} U$& $\mathrm{w}_{\mathrm{defect}}$ & $\mathrm{\omega_0} / 2 \pi$\\
    \hline
    A & $\mathrm{1.65 \, MHz}$ & $\mathrm{3.8 \, fm}$ & 0.55 &$\mathrm{5.6 \times 10^{\minus 7}}$&\SI{15}{\micro \meter} & \SI{189}{\kilo \hertz}\\
    B & $\mathrm{2.55 \, MHz}$ & $\mathrm{1.4 \, fm}$ & 0.17 &$\mathrm{2.3 \times 10^{\minus 2}}$&\SI{130}{\micro \meter} &\SI{174}{\kilo \hertz}\\
    C & $\mathrm{1.70 \, MHz}$ & $\mathrm{2.2 \, fm}$ & 0.19 &$\mathrm{6.9 \times 10^{\minus 2}}$&\SI{30}{\micro \meter} &\SI{242}{\kilo \hertz}\\
    D & $\mathrm{3.41 \, MHz}$ & $\mathrm{0.6 \, fm}$ & 0.24 &$\mathrm{3.2 \times 10^{\minus 2}}$&\SI{30}{\micro \meter} &\SI{246}{\kilo \hertz}\\
    E & $\mathrm{2.39 \, MHz}$ & $\mathrm{1.1 \, fm}$ & 0.55 &$\mathrm{1.9 \times 10^{\minus 8}}$&\SI{80}{\micro \meter} &\SI{181}{\kilo \hertz}\\
    \hline
    \end{tabular}
    \caption{Tabulation of key parameters for each symmetric defect mode for all fabricated devices. Note the correlation between $\mathrm{\Delta}$ and phononic isolation ($\Delta U$).}
    \label{fig:deviceparams}
    \end{figure}
    
    \section{Conclusion}
    To summarize, we studied SiN membrane designs composed of a defect surrounded by a PnC structure. We demonstrated the effects that high contrast PnCs can have on producing wide bandgaps. Additionally, we presented numerical analysis that connects high contrast PnCs to robust energy isolation of defect modes. Conversely, high contrast crystal, compress the mode spatial distribution, induce more bending and increase the internal loss.  However, the increased internal loss is not severe for certain designs, and may for some applications be worth the trade-off for phononic isolation and requirements on the number of unit cells. We have shown that a 1D model with proper conversion, based on the Euler-Bernoulli equation or its reduction to the wave equation, is suitable for analyzing the behavior of a 2D PnC structure. However, incorporating a defect to a 2D crystal requires a complex redistribution of the stress, which could be achieved only by a suitable FEM simulation.

   Utilizing these tools, we designed, fabricated and measured 5 devices differing in PnC contrast and defect designs. These devices exhibit force sensitivity on the order 10 $\mathrm{aN}/\sqrt{\mathrm{Hz}}$ at 4K. We draw attention to both devices A and C, which we predict to have comparable force sensitivities when bending-loss limited. This similar performance is achieved despite the vast difference between the PnCs of the devices and we attribute this similarity in force sensitivity to the low-mass trampoline defect common to both devices. However, under conditions where both these devices are radiation-loss dominated, device A has the potential to outperform device C in force sensitivity due to its robust acoustic isolation, or equivalently achieve equal radiation isolation with a smaller number of unit cells. The low mass trampoline defect archetype also incorporates high stress tethers, which can be readily utilized to deposit metallic or magnetic materials for further device functionalization~\cite{fischer2019spin}.
    
    For force sensing functionalization, it is generally required to add additional components to the device to induce a force between the resonator and the sample in order to couple the mechanical resonator to the force in question. These additional elements add mass and alter the designed mechanical mode shapes and properties. One particularly challenging aspect is how to position the defect mode's frequency within the bandgap upon loading of additional mass. High contrast PnCs offer a larger design space in which to position the defect within the bandgap, while allowing a relatively large tolerance for mass deposition. However,  the low mass defects studied here displayed a first symmetric mode at the lower end of the bandgap. This problem could be adverted by exploring more exotic defects beyond the trampoline archetype, such as delocalized double-pad defects, or by utilizing higher order modes.
    
    As mentioned earlier, the low frequency modes, or notably fundamental mode frequencies ($\omega_0$ in Fig.~\ref{fig:deviceparams}) of a mechanical structure can induce instabilities when placed inside an optical cavity. High contrast PnCs can achieve mode localization and acoustic isolation with fewer unit cells (Fig.~\ref{fig:pnc_design}), and thus have higher fundamental frequencies (Fig.~\ref{fig:deviceparams}). However, the fundamental modes of these high contrast devices have low effective masses when compared to their low-contrast counterparts, and thus have more Brownian motion. Ultimately, the effect on the cavity stability will be determined by the complete mechanical spectrum, which requires knowledge of the dissipation properties of all mechanical modes, and the most relevant effects will be dependent on the optical cavity in question.
    
    All devices reported in this work were fabricated with a thickness of 100 nm. However, by using 20 nm SiN or thinner, force sensitivity could be significantly enhanced thanks to the scaling of $\sqrt{S_{FF}}\propto h^{-3/2}$. Reducing the thickness of device A to 20 nm, would optimally improve its sensitivity to 1 $\mathrm{aN}/\sqrt{\mathrm{Hz}}$ force sensitivity at 4 K. Furthermore, employing SiN resonators below 100 mK, should significantly enhance force sensitivities as $E_2$ of silicon nitride decreases even more under these conditions~\cite{yuan2015silicon}. At these temperatures, structures with low bending loss may become radiative loss limited. In this regime, it may be possible to systematically study the mechanisms of radiative loss in silicon nitride and its transition to internal loss, as has been performed for other mechanical systems~\cite{maccabe2019phononic}. Another consideration for future studies will be the effect of tether size on thermalization of different PnC designs to the cryogenic environment.

    \acknowledgements{We acknowledge funding from AFOSR PECASE, the NSF under grant number PHYS 1734006,  ARO-LPS Cross-Quantum Systems Science \& Technology program (grant W911NF-18-1-0103), and a Cottrell Scholar award. We thank Yeghishe Tsaturyan for helpful discussions.}
    
    \section{References}
    \bibliographystyle{prsty_all}

\begin{thebibliography}{10}

\bibitem{rugar2004single}
D. Rugar, R. Budakian, H. Mamin, and B. Chui, Single spin detection by magnetic
  resonance force microscopy, {\it Nature} {\bf 430},  329  (2004).

\bibitem{poggio2010force-detected}
M. Poggio and C. Degen, Force-detected nuclear magnetic resonance: recent
  advances and future challenges, {\it Nanotechnology} {\bf 21},  342001
  (2010).

\bibitem{safavi-naeini2010proposal}
A.~H. Safavi-Naeini and O. Painter, Proposal for an {Optomechanical}
  {Traveling} {Wave} {Phonon}-{Photon} {Translator}, {\it New Journal of
  Physics} {\bf 13},  32  (2010).

\bibitem{higginbotham2018harnessing}
A. Higginbotham, P. Burns, M. Urmey, R. Peterson, N. Kampel, B. Brubaker, G.
  Smith, K. Lehnert, and C. Regal, Harnessing electro-optic correlations in an
  efficient mechanical converter, {\it Nature Physics} {\bf 14},  1038  (2018).

\bibitem{rabl2010quantum}
P. Rabl, S.~J. Kolkowitz, F.~H.~L. Koppens, J.~G.~E. Harris, P. Zoller, and
  M.~D. Lukin, A quantum spin transducer based on nanoelectromechanical
  resonator arrays, {\it Nat. Phys.} {\bf 6},  602  (2010).

\bibitem{aspelmeyer2014cavity}
M. Aspelmeyer, T.~J. Kippenberg, and F. Marquardt, Cavity optomechanics, {\it
  Rev. Mod. Phys.} {\bf 86},  1391  (2014).

\bibitem{ekinci2005nanoelectromechanical}
K. Ekinci and M. Roukes, Nanoelectromechanical systems, {\it Review of
  scientific instruments} {\bf 76},  061101  (2005).

\bibitem{wilson2008intrinsic}
I. Wilson-Rae, Intrinsic dissipation in nanomechanical resonators due to phonon
  tunneling, {\it Physical Review B} {\bf 77},  245418  (2008).

\bibitem{verbridge2008megahertz}
S.~S. Verbridge, H.~G. Craighead, and J.~M. Parpia, A megahertz nanomechanical
  resonator with room temperature quality factor over a million, {\it Appl.
  Phys. Lett.} {\bf 92},  13112  (2008).

\bibitem{zwickl2008high}
B.~M. Zwickl, W.~E. Shanks, A.~M. Jayich, C. Yang, A.~C.~B. Jayich, J.~D.
  Thompson, and J.~G.~E. Harris, High quality mechanical and optical properties
  of commercial silicon nitride membranes, {\it Appl. Phys. Lett.} {\bf 92},
  103125  (2008).

\bibitem{unterreithmeier2010damping}
Q.~P. Unterreithmeier, T. Faust, and J.~P. Kotthaus, Damping of nanomechanical
  resonators, {\it Phys. Rev. Lett.} {\bf 105},  027205  (2010).

\bibitem{yu2012control}
P.-L. Yu, T. Purdy, and C. Regal, Control of material damping in high-{Q}
  membrane microresonators, {\it Phys. Rev. Lett.} {\bf 108},  083603  (2012).

\bibitem{yuan2015silicon}
M. Yuan, M.~A. Cohen, and G.~A. Steele, Silicon nitride membrane resonators at
  millikelvin temperatures with quality factors exceeding $10^8$, {\it Appl.
  Phys. Lett.} {\bf 107},  263501  (2015).

\bibitem{norte2016mechanical}
R.~A. Norte, J.~P. Moura, and S. Gr{\"o}blacher, Mechanical {Resonators} for
  {Quantum} {Optomechanics} {Experiments} at {Room} {Temperature}, {\it Phys.
  Rev. Lett.} {\bf 116},  147202  (2016).

\bibitem{reinhardt2016ultralow-noise}
C. Reinhardt, T. M{\"u}ller, A. Bourassa, and J.~C. Sankey, Ultralow-{Noise}
  {SiN} {Trampoline} {Resonators} for {Sensing} and {Optomechanics}, {\it Phys.
  Rev. X} {\bf 6},  021001  (2016).

\bibitem{barasheed2016optically}
A.~Z. Barasheed, T. M{\"u}ller, and J.~C. Sankey, Optically defined mechanical
  geometry, {\it Phys. Rev. A} {\bf 93},  053811  (2016).

\bibitem{tsaturyan2017ultracoherent}
Y. Tsaturyan, A. Barg, E.~S. Polzik, and A. Schliesser, Ultracoherent
  nanomechanical resonators via soft clamping and dissipation dilution, {\it
  Nature Nanotech.} {\bf 12},  776  (2017).

\bibitem{ghadimi2018elastic}
A.~H. Ghadimi, S.~A. Fedorov, N.~J. Engelsen, M.~J. Bereyhi, R. Schilling,
  D.~J. Wilson, and T.~J. Kippenberg, Elastic strain engineering for ultralow
  mechanical dissipation, {\it Science} {\bf 360},  764  (2018).

\bibitem{alegre2010quasi-two-dimensional}
T.~P.~M. Alegre, A. Safavi-Naeini, M. Winger, and O. Painter,
  Quasi-two-dimensional optomechanical crystals with a complete phononic
  bandgap, {\it Opt. Express} {\bf 19},  5658  (2010).

\bibitem{yu2014phononic}
P.-L. Yu, K. Cicak, N. Kampel, Y. Tsaturyan, T. Purdy, R. Simmonds, and C.
  Regal, A phononic bandgap shield for high-Q membrane microresonators, {\it
  Appl. Phys. Lett.} {\bf 104},  023510  (2014).

\bibitem{tsaturyan2014demonstration}
Y. Tsaturyan, A. Barg, A. Simonsen, L.~G. Villanueva, S. Schmid, A. Schliesser,
  and E.~S. Polzik, Demonstration of suppressed phonon tunneling losses in
  phononic bandgap shielded membrane resonators for high-{Q} optomechanics,
  {\it Opt. Express} {\bf 22},  6810  (2014).

\bibitem{cole2014tensile}
G.~D. Cole {\it et~al.}, Tensile-strained InxGa1- xP membranes for cavity
  optomechanics, {\it Applied Physics Letters} {\bf 104},  201908  (2014).

\bibitem{buckle2018stress}
M. B{\"u}ckle, V.~C. Hauber, G.~D. Cole, C. G{\"a}rtner, U. Zeimer, J. Grenzer,
  and E.~M. Weig, Stress control of tensile-strained In1- x Ga x P
  nanomechanical string resonators, {\it Applied Physics Letters} {\bf 113},
  201903  (2018).

\bibitem{scozzaro2016magnetic}
N. Scozzaro, W. Ruchotzke, A. Belding, J. Cardellino, E.~C. Blomberg, B.~A.
  McCullian, V.~P. Bhallamudi, D.~V. Pelekhov, and P.~C. Hammel, Magnetic
  resonance force detection using a membrane resonator, {\it Journal of
  Magnetic Resonance} {\bf 271},  15   (2016).

\bibitem{fischer2019spin}
R. Fischer, D.~P. McNally, C. Reetz, G.~G.~T. Assumpcao, T.~R. Knief, Y. Lin,
  and C.~A. Regal, Spin detection with a micromechanical trampoline: {Towards}
  magnetic resonance microscopy harnessing cavity optomechanics, {\it New
  Journal of Physics} {\bf 21},  043049  (2019).

\bibitem{andrews2014bidirectional}
R. Andrews, R. Peterson, T. Purdy, K. Cicak, R. Simmonds, C. Regal, and K.
  Lehnert, Bidirectional and efficient conversion between microwave and optical
  light, {\it Nature Physics} {\bf 10},  321  (2014).

\bibitem{purdy2012cavity}
T.~P. Purdy, R.~W. Peterson, P.-L. Yu, and C.~A. Regal, Cavity optomechanics
  with {Si}₃{N}₄ membranes at cryogenic temperatures, {\it New Journal of
  Physics} {\bf 14},  115021  (2012).

\bibitem{nichol2012nanomechanical}
J.~M. Nichol, E.~R. Hemesath, L.~J. Lauhon, and R. Budakian, Nanomechanical
  detection of nuclear magnetic resonance using a silicon nanowire oscillator,
  {\it Phys. Rev. B} {\bf 85},  054414  (2012).

\bibitem{saulson1990thermal}
P.~R. Saulson, Thermal noise in mechanical experiments, {\it Physical Review D}
  {\bf 42},  2437  (1990).

\bibitem{fedorov2019generalized}
S.~A. Fedorov, N.~J. Engelsen, A.~H. Ghadimi, M.~J. Bereyhi, R. Schilling,
  D.~J. Wilson, and T.~J. Kippenberg, Generalized dissipation dilution in
  strained mechanical resonators, {\it Physical Review B} {\bf 99},  054107
  (2019).

\bibitem{schmid2011damping}
S. Schmid, K. Jensen, K. Nielsen, and A. Boisen, Damping mechanisms in high-{Q}
  micro and nanomechanical string resonators, {\it Phys. Rev. B} {\bf 84},
  165307  (2011).

\bibitem{Villanueva2014evidence}
L.~G. Villanueva and S. Schmid, Evidence of Surface Loss as Ubiquitous Limiting
  Damping Mechanism in SiN Micro- and Nanomechanical Resonators, {\it Phys.
  Rev. Lett.} {\bf 113},  227201  (2014).

\bibitem{maccabe2019phononic}
G.~S. MacCabe, H. Ren, J. Luo, J.~D. Cohen, H. Zhou, A. Sipahigil, M.
  Mirhosseini, and O. Painter, Phononic bandgap nano-acoustic cavity with
  ultralong phonon lifetime, {\it arXiv preprint arXiv:1901.04129}  (2019).

\end{thebibliography}

    \section{Appendix A}
    We present here the method for converting a 2D unit cell geometry to an effective 1D geometry. This approach involves first considering a wave direction $\hat{k}$ along which we will convert the 2D geometry. We wish to encode the geometry transverse to the wave direction into the resulting 1D geometry. It is known that 1D strings with variable width will have lower wave velocities in the wider regions. We imagine that with more transverse\red{?}, the plane wave in the 2D geometry will also have lower wave velocity. Our conversion will then be as follows:
    
    \begin{equation}
        g_{\mathrm{1D}}(x_{\mathrm{l}}) = \int_G g(x_{\mathrm{l}},x_{\mathrm{t}}) dx_{\mathrm{t}}
        \label{eq:1dconvert}
    \end{equation}
    where $x_\mathrm{l}$ and $x_\mathrm{t}$ refer to the coordinates along the longitudinal and transverse wave directions respectively, and $G$ refers to the domain of a single unit cell.
    
    Upon performing the conversion, $g_{\mathrm{1D}}(x)$ gives a continuous variable width of the beam from which a variable wave velocity can be derived. To derive the bandstructure, the 1D wave equation with Floquet boundary conditions can be numerically solved for an arbitrary $g_{\mathrm{1D}}(x)$. However, we find it fruitful to further simplify the geometry where possible. Namely, we wish to convert wave velocity to match that of the Kronig-Penney model which can be solved analytically. This involves defining regions of high mass and low velocity (pads) and low mass and high velocity (tethers) from an arbritrary $g_{\mathrm{1D}}(x)$. 
    
    For unit cells like those in device A, the selection of the pad and tether regions follows naturally from the 2D geometry. Furthermore, the integration carried out in Eq.~\ref{eq:1dconvert} can be carried out using geometric properties. This procedure will perform the transformation $(w_p,w_t,l_p,l_t) \rightarrow (\tilde{w}_p,\tilde{w}_t,\tilde{l}_p,\tilde{l}_t)$ of the geometric parameters.
    
    \begin{figure}
        \centering
        \includegraphics[width=65mm]{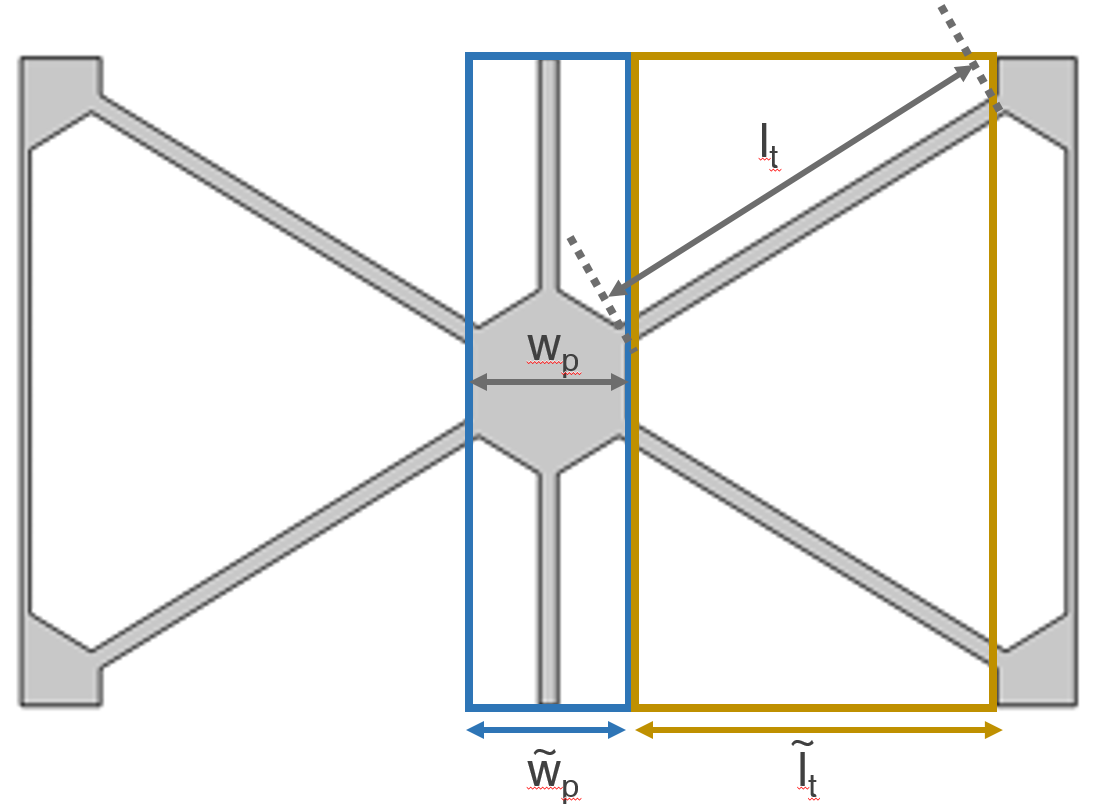}
        \caption{Geometry of the 2D to 1D conversion for a pad-tether unit cell. The blue box (orange box) defines the pad region (tether region) of the PnC unit cell.}
        \label{fig:conversion}
    \end{figure}
    
    Following this procedure, we derive the following:
    
    \begin{equation}
        \rho_p = \frac{\sqrt{3}}{2} w_p h \rho + \frac{l_t t_w h}{w_p}
    \end{equation}
    \begin{equation}
        \rho_t = \frac{4}{\sqrt{3}} w_p h \rho
    \end{equation}
    
    \begin{equation}
        \tilde{l}_p = l_p
    \end{equation}
    
    \begin{equation}
        \tilde{l}_t = \frac{\sqrt{3}}{2} l _t
    \end{equation}
    
    \begin{equation}
        \tilde{\mathcal{T}} = \frac{4}{\sqrt{3}} \sigma t_w h
    \end{equation}
    
    \begin{equation}
        \tilde{v}_p=\big[\frac{3 w_p \rho }{8 t_w \sigma}+\frac{\sqrt{3} l \rho}{4 w_p \sigma}\big]^{\minus \frac{1}{2}}
    \end{equation}
    
    \begin{equation}
        \tilde{v}_t = \sqrt{\frac{\sigma}{\rho}}
    \end{equation}

\begin{figure}
        \centering
        \includegraphics[width=80mm]{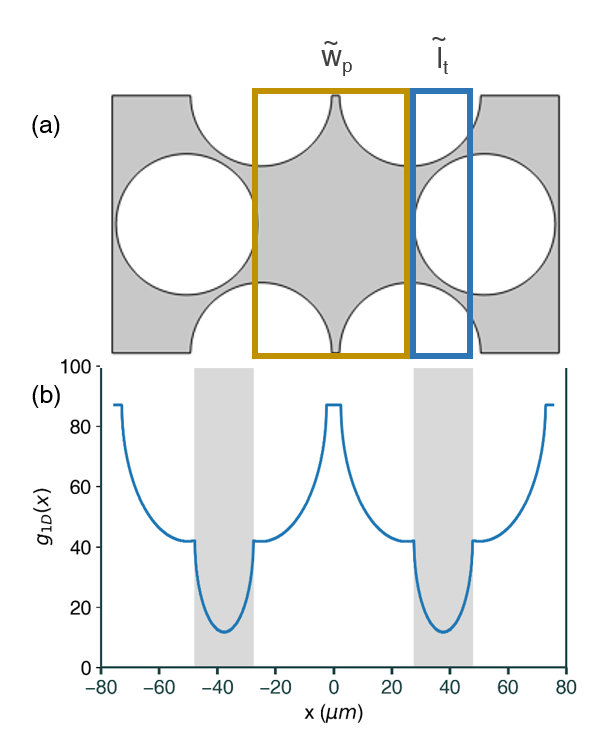}
        \caption{Schematic of 1D collapse for a low contrast unit cell. (a) Definition of pad and tether regions inside the low contrast unit cell. (b) Resulting 1D geometry from performing the collapse. The grey regions indicate the high velocity tether regions used to define $V$ for this unit cell.
        }
        \label{fig:yeghe_collapse}
    \end{figure}

\begin{figure}
        \centering
        \includegraphics[width=80mm]{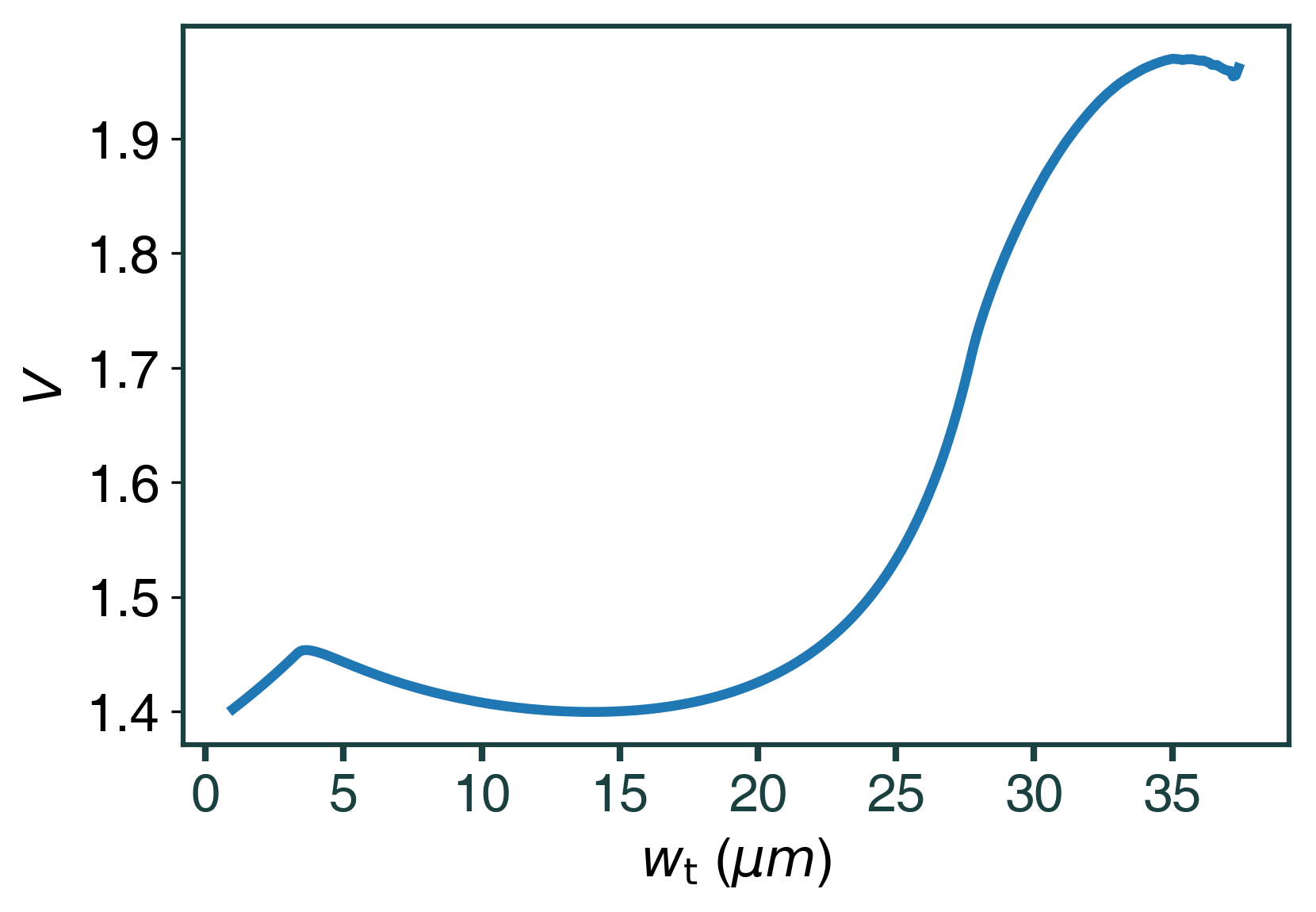}
        \caption{Distribution of contrast values derived from different partitioning of the 1D geometry.
        }
        \label{fig:yeghe_collapse_dist}
    \end{figure}

    In this work we also study unit cells that do not fit the pad-tether model at the outset~\cite{tsaturyan2017ultracoherent}. Fig.~\ref{fig:yeghe_collapse} shows an example collapse in relation to the 2D unit cell. $g_{\mathrm{1D}}(x)$ exhibits pronounced dips over short regions which we will call the tether regions; all other regions will be considered pads. We can then extract the pad width and tether width by taking the mean width over both regions. It is of note that the definition of the pad and tether regions is an arbitrary choice. However, we see that over a reasonable and wide range of tether definitions, $V$ stays roughly constant between 1.4 and 1.5 (Fig.~\ref{fig:yeghe_collapse_dist}). 

    \section{Appendix B}
    \label{sec:appendixB}
    In this section we will present a detailed derivation of bandstructure in the 1D model. We assume that the displacement field has the form
    
    \begin{equation}
        y_k = A_k e^{i k x - i \omega t} + B_k e^{- i k - i \omega t}
    \end{equation}
    
    In this model we view the PnC structure as a modulation in the group velocity of the this acoustic wave. Interfaces between pads and tethers will then have reflection and transmission coefficients given as the familiar Fresnel coefficients
    
    \begin{equation}
        r_{pt} = \frac{v_t - v_p}{v_t + v_p}
    \end{equation}
    
    \begin{equation}
        t_{pt} = \frac{2 v_t}{v_t + v_p}
    \end{equation}
    
    \begin{equation}
        r_{tp} = \frac{v_p-v_t}{v_t+v_p}
    \end{equation}
    
    \begin{equation}
        t_{tp} = \frac{2 v_p}{v_t + v_p}
    \end{equation}
    
    To continue with our analysis, we can derive the motion across as a unit cell as a transfer matrix $M$, which describes the transformation of the amplitude coefficients $A_k$ and $B_k$ across each unit cell of the PnC. The total matrix $M$ will then just be a product of the matrices for each subsection
    
    \begin{equation}
        M = M_{hp} M_{tp} M_{t} M_{pt} M_{hp}
    \end{equation}
    Here $M_{hp}$ captures the accumulated phase of the plane waves across the half pads at either end of the unit cell. $M_t$ does the same for the tether section, while $M_{tp}$ and $M_{pt}$ account for transmission and reflection at each interface. 
    
    \begin{equation}
        M_{hp} = \begin{pmatrix} e^{i k_p l_p / 2} & 0 \\ 0 & e^{- i k_p l_p / 2} \end{pmatrix} 
    \end{equation}
    \begin{equation}
        M_{pt} = \begin{pmatrix} t_{pt} - \frac{r_{tp}r_{pt}}{t_{tp}} & \frac{r_{tp}}{t_{tp} }\\ \minus \frac{r_{pt}}{t_{tp}} & \frac{1}{t_{tp}}
        \end{pmatrix}
    \end{equation}
    \begin{equation}
        M_t = \begin{pmatrix} e^{i k_t l_t} & 0 \\ 0 & e^{-i k_t l_t}
        \end{pmatrix}
    \end{equation}
    \begin{equation}
        M_{tp} = \begin{pmatrix}
        t_{tp} - \frac{r_{pt}r_{tp}}{t_{pt}} & \frac{r_{pt}}{t_{pt}} \\ 
        \minus \frac{r_{tp}}{t_{pt}} & \frac{1}{t_{pt}}
        \end{pmatrix}
    \end{equation}
    
    If we assume that our structure is infinite, then the Bloch condition implies
    
    \begin{equation}
        \begin{pmatrix} A_1 \\ B_1 
        \end{pmatrix} = e^{i K a} \begin{pmatrix} A_0 \\ B_0
        \end{pmatrix}
    \end{equation}
    
    where 
    \begin{equation}
        \begin{pmatrix} A_1 \\ B_1 
        \end{pmatrix} = M \begin{pmatrix} A_0 \\ B_0
        \end{pmatrix}
    \end{equation}
    
    Therefore we have the following eigenvalue problem 
    
    \begin{equation}
        \begin{pmatrix}
        M_{11} & M_{12} \\ 
        M_{21} & M_{22}
        \end{pmatrix}
        \begin{pmatrix}
        A_0 \\ B_0
        \end{pmatrix}
        = e^{i K a} \begin{pmatrix} A_0 \\ B_0
        \end{pmatrix}
    \end{equation}
    
    We note that $M$ is a product of unitary matrices, which places a contraint on its components:
    
    \begin{equation}
        M_{11} M_{22} - M_{12} M_{21} = 1
    \end{equation}
    
    Solving the eigenvalue problem gives us the implicit bandgap equation
    
    \begin{equation}
        e^{i K a} = S \pm \sqrt{S^2 -1}
    \end{equation}
    
    \begin{equation}
        S \equiv \frac{M_{11} + M_{22}}{2}
    \end{equation}
    
    or written more explicitly as:
    
    \begin{equation}
    \begin{split}
    S = \frac{(V+1)^2}{4V}\cos{\omega (t_p + t_t)}\\ - \frac{(V-1)^2}{4V}\cos{\omega (t_p-t_t)}
    \end{split}
    \label{eq:appendixS}
    \end{equation}
    
    \section{Appendix C}
    Devices were fabricated on a \SI{375}{\micro \meter} thick 3 inch diameter silicon wafer with 100 nm of grown stoichiometric LPCVD silicon nitride on either side. Designs were patterned using a direct write photolithography system after spinning \SI{1}{\micro \meter} of SPR-660 photoresist onto either side. The top of the wafer was patterned with the PnC designs, while the back was patterned with rectangular windows aligned to each PnC. The patterning was done on both sides with \SI{300}{\milli \joule \per \centi \meter \squared} of \SI{405}{\nano \meter} light. During these steps, the wafer was affixed to a sapphire carrier wafer with Crystalbond 509 in order to protect the bottom side from unwanted processing. Patterning of the silicon nitirde was completed via a $\mathrm{CF_4}$ reactive ion etch. The wafer was then cleaned with $\mathrm{O_2}$ plasma followed by ultrasound cleaning in an acetone bath. Additional cleaning was performed with isopropyl alcohol and water. To suspend the PnC structures, the window side of the wafer was etched using a 80 C KOH bath. The PnC side was protected via a PEEK wafer holder. Following wet etching, the wafer was cleaned in a Nanostrip bath, acetone and isopropyl alcohol.

    \section{Appendix D}
    FEM simulations in this work were performed using COMSOL Multiphysics. Simulations were performed in 2 steps; the stationary stress redistribution was calculated, followed by a eigenfrequency analysis to determine the mode spectrum of each PnC. The simulations were performed with increasingly fine meshes until both the mode frequencies and loss factors were found to converge to a change less than 5 percent. These simulations assumed that the density of silicon nitride was \SI{3100}{\kilo \gram \per \meter \cubed}, the film stress was \SI{1.13}{\giga \pascal}, the poisson ratio was .27, and the Youngs' modulus was \SI{250}{\giga \pascal}.\\ \\ \\ \\ \\

    \end{document}